# Optical Marking of Alcohol Induced Hemoglobin Modification

R. Vlokh[1], I. Vlokh[2], O. Moroz[3], Yu. Nastishin[1], K. Dudok[4], T. Dudok[1], N. Grinchishin[3], I. Nechiporenko[2], A. Hul[2]

[1]Institute of Physical Optics, 23 Dragomanov St., 79005, Lviv, Ukraine,

[2]Department of Psychiatry, Psychology and Sexology, Lviv National Medical University, 95 Kulparkivska St., 79011, Lviv, Ukraine,

[3]Scientific & Research Laboratory, Lviv National Medical University, Pekarska St., 79010, Lviv, Ukraine,

[4]Biochemical Department, Lviv National University, 4 Hrushevsky St., 79005, Lviv, Ukraine

**Abstract**

It has been shown that conformational modifications of Hb induced by ethanol consumption can be visualized in optical spectra studying oxygenation kinetics of hemoglobin or mixing hemoglobin with Cibacron blue dye. Better dye affinity of blood proteins extracted from alcoholised rats with respect to those from non-alcoholised ones confirms that ethanol and its metabolites induce structural pathologies in blood protein molecules. The detected changes for the case of the posterity of intoxicated animals may be explained as a post-translation modification, as well as a disturbance of the structure and function of tissue cellular gene mechanism for the blood creation. It is established that alcohol intake during first four months leads to the decrease of fractional weight of oxyhemoglobin and to the increase of methemoglobin amount in blood. Further alcohol consumption is accompanied by recovering of the normal level of hemoglobin derivatives in blood. Normalization of the fractional weight of hemoglobin derivatives in blood after durable (longer than 5-6 months) ethanol intoxication is most probably due to the activation of the enzyme (acetaldehyde dehydrogenase) system, lowering the level of acetaldehydes in blood.

# 1. INTRODUCTION

Classification of alcoholism as a mental disease, including its genetic etiology, is a dominating conception for this pathology in current medical literature [1]. Such an approach is quite perspective because it implies that the predisposition to this pathology can be detected, its prophylaxis and early diagnosis suppose to be possible and finally that as a disease it can be treated. Understanding of the origin of clinical characteristics of alcoholism as well as other mental diseases is complicated by the inaccessibility of the cerebrum as the main investigation object. One of the main directions of the development of non-invasive techniques for diagnosis of mental diseases traditionally is the examination of biological liquids (blood, urine, liquor, *etc.*) [2]. There are numerous biochemical and biophysical characterization techniques for the detection of pathological features accompanying a disease, which can be searched in the properties of biological liquids or their components. In many cases these features are related to the presence of a substance, so-called ***biomarker***, unambiguously pointing to the disease. The detection of biomarkers is based on their specific biochemical properties: characteristic chemical reactions, special response to external actions [3,4] etc. and, thus, in a wider sense a biomarker can be interpreted as a specific property of the investigated biological substance, unambiguously indicating a pathology. Such markers can be classified as ***biochemical markers***. Therefore, under ***biochemical*** markers one understands those features caused by a disease and consisting in deviations in chemical composition and/or concentration of the biological liquids, and which in limiting cases reduce to the absence/presence of some composition components (substances or bodies) or in the disturbances of their biochemical activity. However, it is clear that according to the definition given above the biochemical markers do not cover all the features, which can be classified as markers of a disease, because there are those that, strictly speaking, do not imply changes in chemical composition, but consist in the modifications of the structural organization of the complex components of biological liquids, such as conformational modifications of biological molecules (in particular in proteins), in pathological changes in the cell membranes, including possible phase transitions, modifications of their surface curvature, changes in their permeability *etc.* (review on physical aspects of cell membrane structure can be found in [5]); we call such features ***structural markers***.

The investigation on the *biochemical markers* of mental diseases and particularly of alcoholism is represented in the literature by numerous publications (see [6-12]). For example, it was found that the weight of protein fractions of blood serum at the ethanol intoxication is changed in comparison with that in the absence of alcohol intoxication. Depression of hemocoagulation [10], the increase of circulated blood volume [11], the decrease in the number of erythrocytes, the increase of the hemoglobin amount and the increase of the amount of fetal hemoglobin [12] have also been identified as characteristic features accompanying alcoholism. They can be classified as *biochemical markers of alcoholism*. **Biochemical markers** of alcoholism are divided into two groups: *markers of the pathological process* **(state markers)** [13,14] and *markers of liability to the disease* **(trait markers)** [15]**.** Lack of activity of ferment Alcohol dehydrogenases is an example of a *trait marker*. Lack of this ferment in the organism leads to the enhancement of acetaldehyde concentration in the organism, that is accompanied by unpleasant feelings, which play a repelling role and form a distaste to the excessive alcohol drinking or even induces total abstinence.

Concerning the *structural markers* the information in the literature is rather sporadic [16]. The scarcity of investigations in this direction can be mainly related to difficulties in the sample preparation. Indeed, most of characterization techniques, that provide structural information, request uniform anisotropic samples, in particular of high optical quality for optical characterization. There are very few techniques for preparation of anisotropic biological samples (see for example [17-21]).

In some cases structural information can be obtained using powdered, unaligned or liquid isotropic samples. Optical investigations of isotropic solutions of biologic materials are capable to bring structural information for biologic macromolecules. In the sequel we call *optical markers* those *biochemical* and *structural markers* that are detected with optical techniques. The importance of the detection of *optical markers* and understanding their origin on the molecular level in relation to the clinical picture of a disease is hard to overestimate accounting their non-invasive character especially for early diagnostics, including the diagnostics of latent forms of a disease.

There are several findings in current literature, which can serve as basic ideas behind the optical detection of biological markers of alcoholism. One of the symptoms accompanying alcoholism is hypoxia. From molecular point of view this pathology is related to the presence of acetaldehydes in biological liquids. Acetaldehydes are products of ethanol oxygenation reactions. The acetaldehyde can interact with proteins and enzymes and induces their modification. Modifications of non-enzyme proteins consist in conformational changes of their molecular structure and, as a result lead to a disturbance of their biological functions. In particular, it is known that acetaldehyde can create complexes with the blood serum albumin; modifies the iron transferring-mediator and hemoglobin-oxygen-transportation proteins [22]. In addition acetaldehyde can have indirect effects, namely, stimulating the peroxide lipid oxidation. Referring to peroxide lipid oxidation, we have in mind the increased level of active metabolites, such as mulonic dialdehyde, as well as the other aldehydes and cetons that manifest cytotoxic influence and are actively modifying agents [7]. These processes induce hyper-production of oxygen in the mitochondria and creation of strong oxidizer, the super oxide anion ($O_2^-$). The latter participates in oxygenation of fatty acids, the components of phospholipids. Thus, entering into the organism, ethanol causes, directly or indirectly, a wealth of secondary metabolites, which induce a disturbance of cytoskeletons of cells, membranes of cell organelles, modify biologically important macromolecules and change their functions [8]. Chemical and structural modifications of blood proteins can be detected via modifications in their optical spectra with respect to the corresponding spectra for proteins in the absence of pathology.

In this paper we focus on comparative spectral investigations of isotropic water solutions of hemoglobin samples extracted from rats, protractedly intoxicated by ethanol, their get and from rats free of ethanol intoxication. The aim of this study is detection of optical markers in hemoglobin molecules induced by ethanol consumption.

## 2. SPECTRA OF HEMOGLOBIN FROM RATS SUBJECTED TO DURABLE ETHANOL INTOXICATION

### 2.1. EXPERIMENTAL

Females of white rats selected with an average weight of 250g have been placed into single cages and separated in two groups, namely: Alcohol Intoxicated group ($AI_{i,j}$, where $i$ denotes the duration of alcohol intoxication in months and $j = 0,1,2,3$ indicates the generation with $j = 0$ corresponding to the parents or to the animals whose get has been not investigated) with free access to 15% water solution of ethanol instead of water and Alcohol Free (AF) group without ethanol intoxication (consumption of water without ethanol).

The rats were copulated in the age of four months after one month of ethanol feeding. Females have continued to intake ethanol during gestation. Amount of consumed liquid and the weight of animals were daily controlled. A parameter $a_m$ calculated as the ratio of the mass of consumed ethanol over the mass of a given animal and characterizing alcohol motivation has been determined for each animal. If $a_m > 7\ g/kg$ we conclude that there is alcohol motivation. In 10 days rats with evident alcohol motivation were selected for further forcible alcohol intoxication. The blood sampling from the tail vein for further investigation was performed after $i$ months of 15% ethanol consumption. For comparison the blood from AF group of animals without ethanol intoxication was similarly studied.

To prepare hemolisate we have used heparin as anticoagulant. Plasma was separated centrifuging whole heparinizated blood. Hemolisation of red blood cells was performed with 30mM K-, Na-phosphate buffer (pH 7.36).

Hemoglobin was extracted from blood following the technique described in (see e.g. [12] for details). The separated hemoglobin was transformed into cyanmethemoglobin (CNMetHb) and methemoglobin (MetHb); their concentrations were determined measuring their light absorption.

To detect conformation modifications of proteins one uses ability of some dye molecules to bound protein molecules. The conformation modifications of protein molecules change their affinity to the dye molecules, *i.e.* the number of the dye molecules attached to the protein molecules depends on the conformation state of the protein molecule. The idea behind is that the spectra of the proteins with dye molecules attached to them differ from those, which are obtained when the dye molecules do not bind the protein molecules. Comparing spectra for dye doped water solutions of hemoglobin extracted from blood taken from animals of $AI_{i,j}$ and AF groups we expect to detect the difference in the conformation states of hemoglobin for these groups. To study the conformational modifications in hemoglobin (Hb) following the approach described in reference [12] we use dye: Cibacron Blue (8mg of cibacron blue dissolved in 100ml of 0.1M acetate buffer with pH=4.8). Optical spectra were measured using spectrometer Specord M-40 in the spectral range $450 - 750\,nm$.

Stability of erythrocyte membranes was studied with the method of acid erythrograms. Erythrocytes were separated from the whole blood plasma by the centrifugation (2000 rpm) during five minutes. The erythrocyte suspension was wetted three times with 0.9% solution of NaCl. The hemolysation was done at 0.004 HCl. Erythrograms were plotted using the results of measurements obtained with photoelectrical calorimeter KFK-3.

## 2.2. ABSORPTION SPECTRA OF HEMOGLOBIN

*Oxy- vs. deoxy-hemoglobine from $AI_{1,0}$ and AF groups.* Depending on the method of preparation the extracted hemoglobin can be in chemically different forms. Light absorbing spectra of deoxygenated (deoxy-Hb, other names: unsaturated or reduced) hemoglobin and oxyhemoglobin (oxy-Hb) differ and this allows one to estimate their respective amount in the sample.

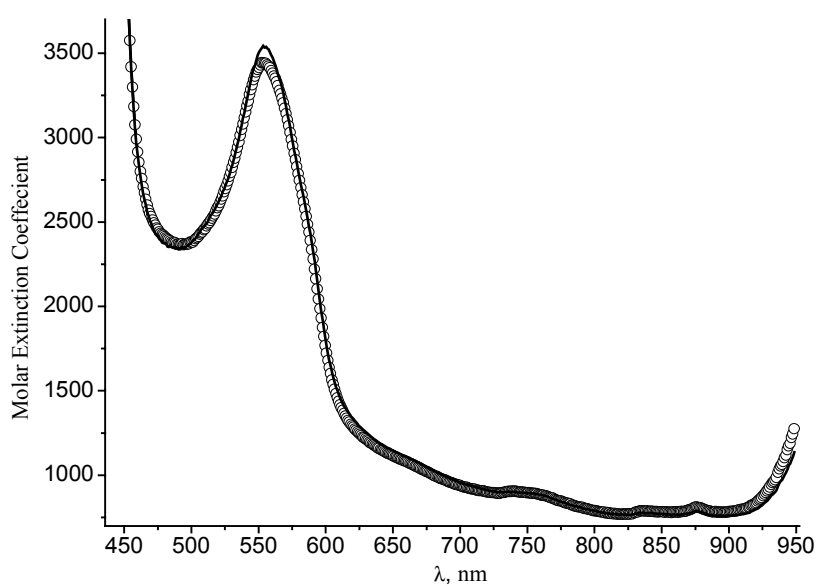

Figure 1. Absorption spectra of hemoglobin from alcohol intoxicated (AI1,0 group: solid curve) and non-intoxicated (AF group: open circles) white rats.

Figure 1 shows absorption spectra for $AI_{1,0}$ and *AF* groups, namely dispersion of the molar extinction coefficient *e* of hemoglobin calculated as:

$$e = A\frac{M}{cd}, \qquad (1)$$

where $A = -\log_{10}(\frac{I}{I_0})$ is the light absorbance of the sample with the light intensities entering and exiting from the sample here denoted as $I_0$ and *I*, respectively; $M = 64500\frac{g}{mole}$ is the molar mass of hemoglobin; *c* in units $[\frac{g}{l}]$ is the concentration of hemoglobin and *d* is the light pass in the sample. According to the literature data [23] presence of a broad absorption band with the maximum at $\lambda = 556\,nm$ is a characteristic feature of the deoxy-form of hemoglobin. Oxy-hemoglobin displays two characteristic absorption maximums at $\lambda_1 = 542\,nm$, $\lambda_2 = 576\,nm$ and minimum at $\lambda = 560\,nm$. Light absorption coefficient measured at $\lambda = 560\,nm$ can be used for monitoring of the deoxygenation kinetics. Qualitatively spectra are similar for all the studied samples of $AI_{i,j}$ and *AF* groups. However they differ quantitatively.

The two plots in Figure 1 have two isobestic points, namely at 590 and $805\,nm$, confirming that concentrations of hemoglobin in both *AI* and *AF* samples are equal. The absorption value at the wavelength $\lambda = 560\,nm$ is higher for $AI_{i,j}$ groups than it is for the *AF* group. This latter suggests that the vein blood of the ethanol intoxicated rats is enriched by deoxy-Hb form in comparison with the group *AF* free from ethanol intoxication. We check this statement comparing kinetics of hemoglobin deoxygenation for alcoholised and non-alcoholised rats.

*Deoxygenation kinetics for $AI_{1,0}$ and AF groups.* Deoxygenation of hemoglobin is induced following special procedure based on reduced pressure using modified saturator. Below we demonstrate that the kinetics of deoxygenation is different for the $AI_{i,j}$ and *AF* groups. We relate this difference to the action of ethanol on the structure of the hemoglobin molecule. The time dependencies of Hb deoxygenation rate have been studied by means of spectroscopic absorption technique at the wavelength 560 nm.

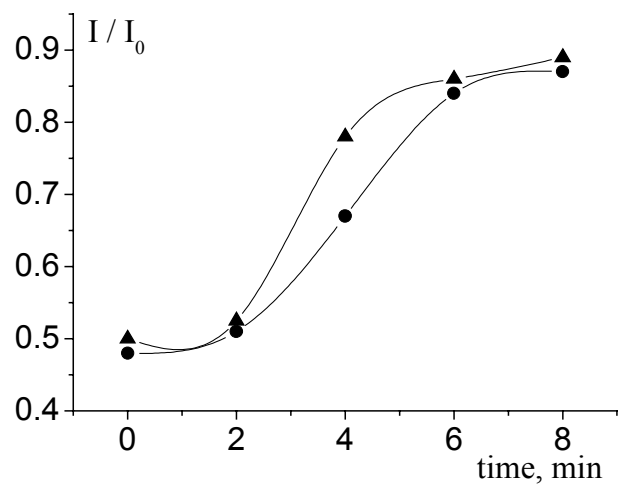

Figure 2. Kinetics of deoxygenating for hemolisate extracted from rats of A1,0 (triangles) and AF control groups (circles).

In due course of the deoxygenation procedure the light absorption at $\lambda = 560$ nm increases and its relative increase is evident in Figure 2, where the ratio $I(t)/I_0$ of the adsorption at a given moment $t$, counted from the beginning of the deoxygenation procedure, over that for $t = 0$ is shown. It is seen that the rate of deoxygenation is higher for hemoglobin extracted from $AI_{1,0}$ group than it is for AF group. In Figure 3 we plot the rates of deoxigenation calculated as time derivatives of the dependencies $I_t(t)/I_0$ in due course of deoxigenation process. For the $AI_{1,0}$ group the rate of deoxygenation reaches its maximal value at $t = 4\min$, while for the $AF$ group the maximal rate of deoxygenation is at 5.5 minute.

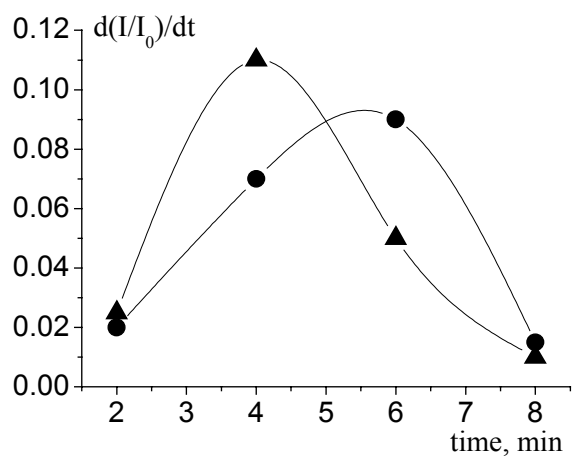

Figure 3. Time dependencies of deoxygenation velocity (plotted using data shown in Figure 2) for hemolisate extracted from alcoholized rats, group $AI_{1,0}$: dots and from rats free of ethanol intoxication, AF control group: open circles.

We have also measured the relative fractional weight of hemoglobin derivatives, namely of: desoxy-, oxy-, carboxy-, sulf- and meth-forms of hemoglobin in hemolysate as a function of the duration of ethanol intoxication of white rats and their posterity.

*Fractional weight of hemoglobin forms in hemolisate determined from their light absorption.* Even slight deviations from the normal level of hemoglobin derivatives have detrimental effects to the tissues. Hemoglobin derivatives except the oxy-form do not permit the transfer of oxygen. Light absorption of whole blood hemolysate results from the superposition of the light absorption by five composition components, derivatives of hemoglobin, namely by oxy-, desoxy-, carboxy-, sulf- and meth-(MetHb) hemoglobin. Each of these five components when dissolved in water as a single solute exhibits its own absorption band. In hemolisate these components are mixed and one can express the resulting light absorption spectrum $A(\lambda)$ as a sum of absorptions of single component multiplied by their concentration in the hemolisate mixture:

$$A(\lambda) = \sum_{i=1}^{5} c_i A_i(\lambda),  \qquad (2)$$

where $c_i$ are molar concentrations of the components, normalized by their total molar mass such that $\sum_{i=1}^{5} c_i = 1$. In hemolisate the hemoglobin forms can be mutually transformed and consequently their concentration are changed at the expense of each other. Sensitivity of the spectral test to the blood composition is well illustrated by the fact that a decrease of the amount of oxyhemoglobin leads to visible change in the blood color from bright red to purple-blue. To characterize hemolisate composition, we introduce a parameter called fractional absorbing weight $v_i$ of a given light-absorbing component calculated as a ratio

$$v_i = \frac{A_i}{\sum_{i=1}^{5} A_i} \qquad (3)$$

of light absorption $A_i$ of hemolisate measured at a given wavelength $\lambda_i$ to the sum of absorption values measured at these five wavelengths $\lambda_i$. By definition we have $\sum_{i=1}^{5} v_i = 1$. The wavelengths $\lambda_i$ are chosen to achieve the condition $c_i \approx v_i$ for each of five studied hemoglobin derivatives forms. In Table I we present our results on measurements of $v_i$ for different groups of rats.

Table I. Effect of alcohol intoxication on fractional absorbing weight $v_i$ (%) of hemoglobin derivatives.

| Grou | Num- | Hemoglobin derivatives |
|---|---|---|

| p | ber of rats | desoxy-Hb | oxy-Hb | carboxy-Hb | sulf-Hb | Met-Hb |
|---|---|---|---|---|---|---|
| $AF$ | 6 | 0.01±0.005 | 93.24±0.2.47 | 2.06±1.12 | 3.22±3.03 | 1.47±1.75 |
| $A_{1,0}$ | 3 | 0.01±0.004 | 87.14±1,47 | 2.66±1.05 | 5.42±0.49 | 4.76±0.23 |
| $A_{2,0}$ | 7 | 0.01±0,004 | 88.35±2,70 | 2.78±0.77 | 4.20±0.81 | 4.88±0.89 |
| $A_{4,0}$ | 4 | 0.01±0.003 | 90.87±1,12 | 3.33±0.35 | 2.30±0.55 | 3.46±0.50 |
| $A_{5,0}$ | 3 | 0.01±0.005 | 97.08±0,82 | 2.10±0.07 | 0.01±0.01 | 0.81±0.80 |
| $A_{6,0}$ | 3 | 0.025±0.005 | 92.66±2.92 | 3.4±0.19 | 3.28±1.21 | 1.65±1.40 |
| $A_{6,1}$ | 6 | 0.01±0.005 | 93.15±2.33 | 4.14±0.22 | 1.45±1.36 | 1.06±0.21 |

According to Table I consumption of ethanol during first 4 months leads to the decrease of the oxyhemoglobin and to the increase of methemoglobin amount in the blood of rats. After 5-6-month of alcohol intake, the values of these parameters return to their normal level. Thus, one can conclude that ethanol affects oxidation of hem iron decreasing fraction of oxyhemoglobin in blood. Normalization of the fractional weights of hemoglobin derivatives at further ethanol intake and even for the next generation of rats can be associated with the activation of enzyme systems (acetaldehyde dehydrogenase, etc.), participating in the decrease the level of acetaldehyde. It is, probably, at this stage that the mechanism of adaptation to the alcohol intake activates.

Table II. The erythrogram parameters of blood for of different groups of rats.

| Groups of animals according to durability of alcohol consumption | Number of rats | Maximum hemolysation, min | Total hemolysation, min | Maximum hemolysation, % |
|---|---|---|---|---|
| | | | | |
| *AF*, control group | 6 | 5.1+0.1 | 8.1 ±0.2 | 38,52±1,90 |
| $A_{1,0}$ | 5 | 5.1+0.2 | 9.4±0.6 | 36,05±3.10 |
| $A_{2,0}$ | 3 | 5.5±0.5 | 9.7±0.7 | 31.94±3.50 |
| $A_{3,0}$ | 3 | 5.0±0.1 | 8.2±1.3 | 37.30±3.70 |
| $A_{4,0}$ | 3 | 4.5 ±0.2 | 7,3 ±0.7 | 39.06±1,30 |
| $A_{5,0}$ | 3 | 2.3 ±0.2 | 6,5 ±0.7 | 35.71 ±0.80 |
| $A_{6,0}$ | 4 | 4.6±0.3 | 8.9±0.6 | 21.37±4.0 |
| $A_{4,1}^{1}$ | 6 | 3.25±0.10 | 8.9±0.4 | 20.90±2.4 |
| $A_{4,1}^{2}$ | 3 | 3.40±0.20 | 7.7±0.5 | 24.09±3.3 |
| $A_{6,1}^{3}$ | 6 | 3.2±0.2 | 9.00±0.7 | 18.83±2.2 |

It is known that the alcohol intoxication is accompanied by the activation of free-radical lipid oxygenation [22]. Free-radical lipids are components of erythrocyte membranes. Their state can be tested using the acid erythrogram method. Our data show that the maximum hemolysation of blood for the control group of rats occurs at $5.1 \pm 0.1 \min$ (see Table II and Figure 4). The corresponding time parameter for the group of rats that consumed ethanol during 1-3 months is in the same range, while the duration of total hemolysation is somewhat longer. For rats that consumed ethanol during 4-6 months, the time of maximum hemolysation and the time of total hemolysation are significantly lowered (from $4.5 \pm 0.2 \min$ to $2.3 \pm 0.2 \min$). For first generation of rats that consumed alcohol during 4-6 months maximum hemolisation is at $3.2 \pm 0.2 \min$, while the time of total hemolisation is $8.4 - 9.8 \min$.

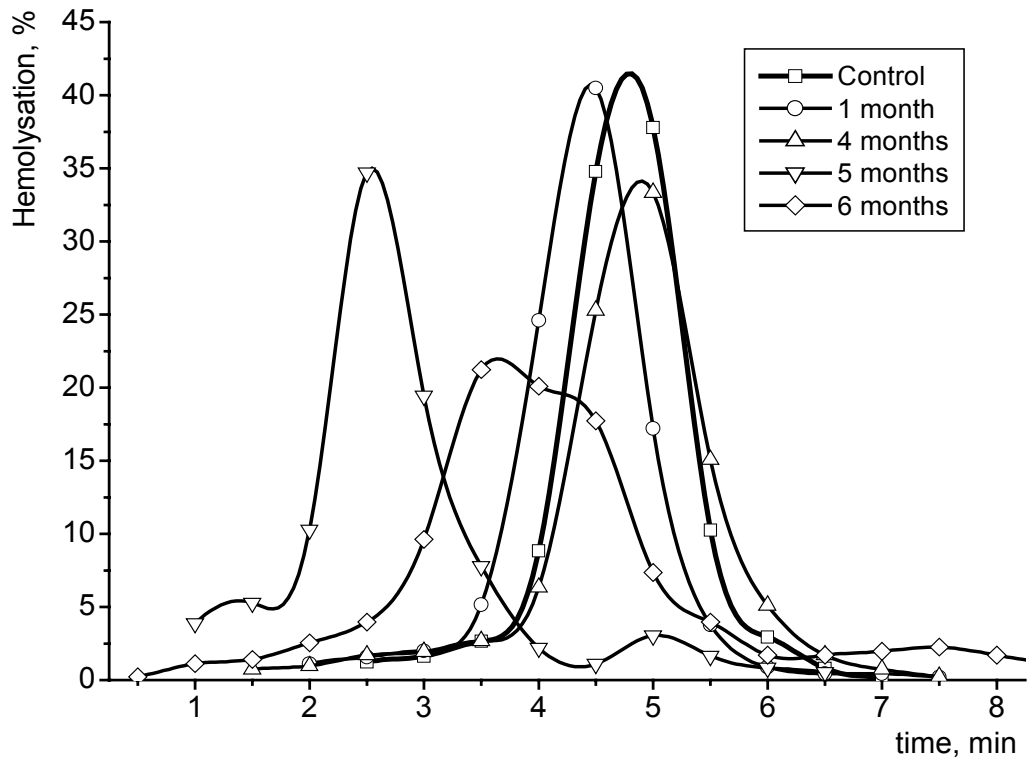

Figure 4. Time dependencies of deoxygenation velocity (plotted using data shown in Fig.2) for hemolisate extracted from alcoholized rats, group Al1,0: triangles and from rats free of ethanol intoxication, AF control group: circles.

Results presented above show that the durable ethanol intake leads to the decrease in the erythrocyte life duration and to the increase in the inhomogeneity of their population and, hence, that the alcohol intoxication induces a disturbances in both structural and functional states of erythrocytes, as well as in the oxygen-transportation function of hemoglobin.

*2.3. ABSORPTION SPECTRA FOR HEMOGLOBIN SOLUTION DOPED WITH CIBACRON BLUE*

Binding of ligands is a property of protein molecules that definitely depends on their conformational state. If a ligand is a light absorbing material (a dye), then affinity of the protein molecules to dye molecules can be characterized studying absorption spectra of the protein solutions doped with dye. Because hemoglobin itself is a light absorbing material its chemical interaction with non-light-absorbing ligands can also be visualized in absorption spectra of hemoglobin samples. Therefore, adding dyes to hemoglobin samples one can visualize structural differences in protein molecules extracted either from alcohol intoxicated $AI_{i,j}$ or non-intoxicated $AF$ groups of rats studying absorption spectra.

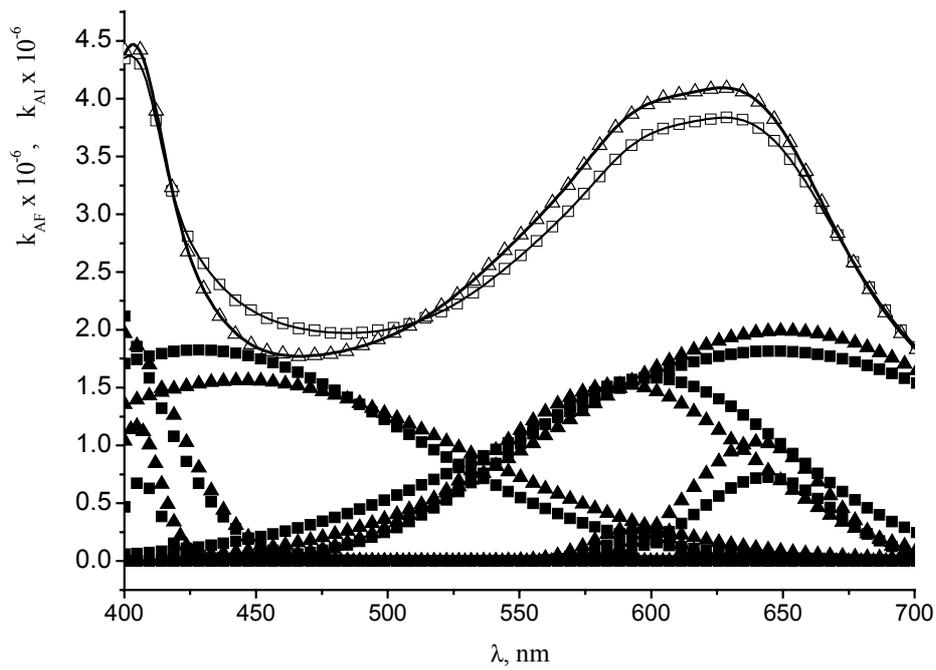

a

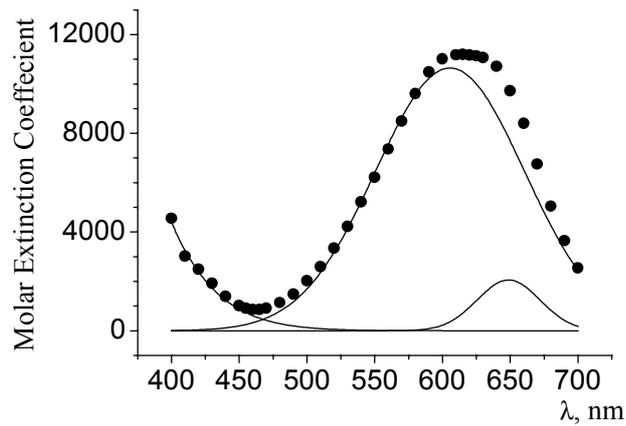

b

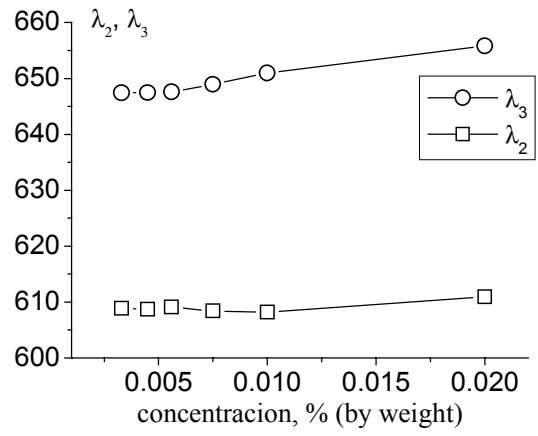

c

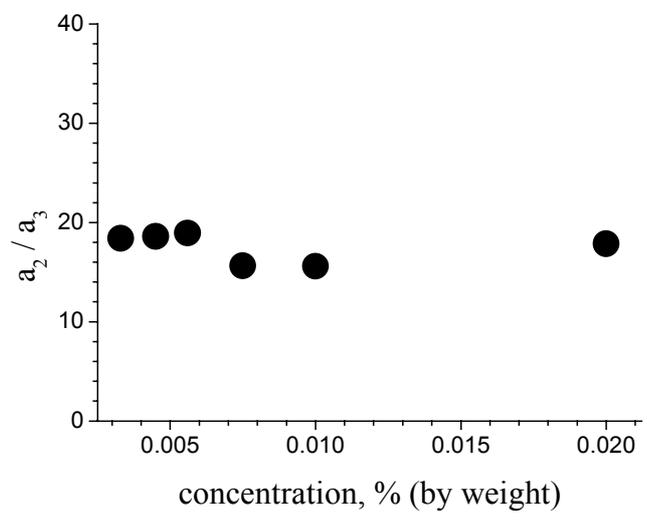

d

Figure 5. (a) Absorption spectra of hemoglobin solution mixed with Cibacron Blue dye for: AI1,0 (open triangles) and AF control (open squares) groups. Fitting of the experimental data by the superposition of seven Gauss-shaped peaks is shown by solid triangless and squares superimposed on the experimental data, respectivelly for AI1,0 and AF groups. (b) Absorption spectra for 0.01 wt.%, (full circles) of Cibacron Blue dye solutions in water. Fitting with three Gaussian peaks (fitting parameters are given in Table 4) is shown by a thin lines. c) Positions of the constitutive peaks λ2 and λ3 . d) Ratio of the fitting absorption weights a2/a3 for Cibacron Blue water solutions for different concentrations.

Dye Cibacron Blue F3G-A was purchased from Fluca. In Figure 5 we compare absorption spectra of hemoglobin water solutions doped by Cibacron Blue F3G-A, a dye known to bind proteins [24-27]. Strong binding of albumin to Cibacron Blue resin column is a base for the most commonly used method for albumin removal from hemolisate. In Figure 5 we represent absorption spectra for the mixtures of hemoglobin extracted from *AF* and *AI$_{1,0}$* groups with dye. The *AF* and *AI* hemoglobin-dye samples where prepared in the equal concentration proportions. Notice that because we deal with a mixture it is more properly to characterize the absorption spectra in the units of the absorption coefficient $k(\lambda)$, which does not require the molar mass of the solute instead of that in units of molar extinction coefficient:

$$k(\lambda) = -\frac{\lambda}{4\pi d} \ln \frac{I}{I_0}. \qquad (4)$$

It is seen from Figure 5,a that in the ranges of short and long wavelength the values of $k_{AF}$ and $k_{AI}$ practically coincide. For short wavelength region (400-430nm) the light absorption is mostly due to hemoglobin molecules (the hemoglobin-to-dye ratio of the molar extinction coefficients is about 50) and, thus, the coincidence of the absorption coefficients $k_{AF}$ and $k_{AI}$ in this spectral region is due to the same number of hemoglobin molecules in both samples, that is agree with the nominally equivalent concentration proportions for the samples, mentioned above. Plots in Figure 5,a corresponding to *AF* and *AI* samples has three isobestic points: at 425, 515 and 675nm. At long wavelength the absorption of hemoglobin drops such that in the region of the absorption band of Cibacron blue ($550\,\text{nm} < \lambda < 675\,\text{nm}$) the light absorption by dye molecules prevails (the hemoglobin-to-dye ratio of the molar extinction coefficients is about 0.3) and, hence, the coincidence between $k_{AF}$ and $k_{AI}$ at long wavelengths reveals roughly the same number of dye molecules in the *AF* and *AI* samples in a accordance with nominally equivalent compositions. Differences between $k_{AF}$ and $k_{AI}$ spectra are observed in the middle part of the studied spectral region, namely: $k_{AF} > k_{AI}$ at $430\,\text{nm} < \lambda < 520\,\text{nm}$, while it is opposite at $520\,\text{nm} < \lambda < 675\,\text{nm}$, where $k_{AF} < k_{AI}$. Characteristic hemoglobin absorption band corresponding to the light absorption by iron centers with the maximum at $\lambda = 556\,\text{nm}$ is not visible in the spectra of both *AF* and *AI* samples (compare Figure1 and Figure 5,a). This latter is an indication that the hemoglobin molecules interact with dye molecules, creating hemoglobin-dye molecular complexes. Indeed, negatively charged Cibacron Blue dye residues can be electrostatically bound to positively charged side groups of hemoglobin, including the iron centers, and shifting their absorption to longer wave lengths and most probably overlapping with the absorption bands due to polyaromatic central cores of the Cibacron dye molecules.

Observation of the sample cuvette with naked eyes shows that the hemoglobin-dye solution is not fully homogeneous, but contains suspended flakes, which in time fall down to the bottom of the cuvette. Presence of these flakes evidences formation of hemoglobin-dye complexes. To assure that the sedimentation of these flakes does not affect measured spectra, the sample was stirred before each experiment; spectral measurements were performed in the direction from shorter to longer wavelengths and then the sample was again stirred and the spectrum was taken in the reverse direction from longer to shorter wavelengths. Coincidence of the spectra taken in two directions excludes influence of the sedimentation from consideration.

Additional precaution to the measurement procedure concerns the property of dyes to adsorb on solid surfaces including glass walls of the measured cuvette. After few days a solid bluish layer covering the cuvette walls can be seen. To avoid this problem all the spectral measurements were performed on freshly prepared samples.

Another evidence for the hemoglobin-dye complexes follows from the comparison of the shapes of absorption bands for Cibacron Blue water solutions and that for their mixtures with hemoglobin. The absorption band for the hemoglobin-dye mixtures (Figure 5,a) is about 20nm broader than it is for pure water dye solutions without hemoglobin (Figure 5,b). The shapes of the absorption bands in Figure 5,a are non-Gaussian and, thus, implying that these bands are composed of at least two constitutive Gaussian bands centered presumably around 615 and 650nm respectively. Notice that the absorption maximum for pure water solutions of Cibacron Blue is located roughly at 615nm and a peak with $\lambda = 650$ nm is not visible, while it shows up as a constitutive peak for the dye mixture with hemoglobin. Thus, to characterize quantitatively the difference in the absorption spectra for *AF* and *AI$_{1,0}$* samples it is worth to make the deconvolution of the spectra into constitutive peaks by fitting both dependences with a function, which is a superposition of Gaussian peaks:

$$k(\lambda) = \sum_{i=1}^{n} \frac{a_i \sqrt{2}}{w_i \sqrt{\pi}} e^{-2\left(\frac{\lambda-\lambda_i}{w_i}\right)^2}, \qquad (5)$$

where $w_i$ are half-widths of the constitutive peaks centered in the absorption maximum at the wavelength $\lambda_i$ with absorption weights $a_i$, a constant proportional to the number of the molecules responsible for the absorption in the given peak. We have started the fitting of the experimental data by Eq.5 with five initial parameters for constitutive Gaussian peaks $\lambda_i$, among which three peaks have been chosen at $\lambda_1^0 = 390\,\text{nm}$, $\lambda_{1a}^0 = 430\,\text{nm}$, $\lambda_{1b}^0 = 556\,\text{nm}$ corresponding to the characteristic hemoglobin peaks and other two peaks: at $\lambda_2^0 = 615\,\text{nm nm}$ and $\lambda_3^0 = 650\,\text{nm nm}$. Last two peaks, namely $\lambda_2^0$ and $\lambda_3^0$ are in the range of the absorption band for Cibacron Blue molecules and thus they can be understood as guess positions for the constitutive peaks composing the absorption band of Cibacron Blue in water. The resulted 5 peaks fitting curves roughly superimposes with the experimental data, though the coincidence is not total (we do not show these graphs here). To improve the fitting we introduced 6$^{th}$ peak at $\lambda_{2a}^0$ between $\lambda_{1b}^0$ and $\lambda_2^0$ peaks and then 7$^{th}$ peak at $\lambda_{3a}^0$ between $\lambda_2^0$ and $\lambda_3^0$ peaks. Insertion of these additional peaks can be approved assigning them to the light absorption of hemoglobin-dye molecular complexes. The best fits of the experimental data in the studied spectral region ($400 \div 700\,\text{nm}$) for 7 constitutive peaks are shown in Figure 5,a by bright lines superimposed on dark symbols plotting experimental data. Fitting results are summarized in Table III. The fitting curves perfectly coincide with experimental $k(\lambda)$ dependencies. It is worth noticing that the variation of the initial guess values for $\lambda_{2a}^0$ and $\lambda_{3a}^0$ peaks within the ranges between $\lambda_{1b}^0$ and $\lambda_2^0$, $\lambda_2^0$ and $\lambda_3^0$ peaks does not change the fitting results. It is important to stress again that to get the fitting results not worse than that shown in Figure 5,a one needs at least 7 constitutive peaks. Although playing with initial guess values for fitting parameters choosing them incidentally one can come to other slightly different sets of $\lambda_i$ values for the fitting parameters (in the worse case we find ±5 nm from those given in Table III), such a random choice of the initial guess values does not seem to be motivated. We, thus, are led to use the fitting parameters mentioned above for the interpretation of the origin of the difference between $k_{AF}$ and $k_{AI}$ spectra.

Table III. Fitting results of absorption spectra of hemoglobin solution mixed with Cibacron Blue dye (Figure 5,a).

| Peak number | Fitting parameter | AF group | $AI_{1,0}$ group |
|---|---|---|---|
| 1 | $\lambda_1$, nm | 374.98 | 395.60 |
| | $a_1$ | $2.28 \times 10^{-4}$ | $1.22 \times 10^{-4}$ |
| | $w_1$ | 62.05 | 48.43 |
| 2 | $\lambda_{1a}$, nm | 406.43 | 404.52 |
| | $a_{1a}$ | $1.24 \times 10^{-5}$ | $2.67 \times 10^{-5}$ |
| | $w_{1a}$ | 14.40 | 18.28 |
| 3 | $\lambda_{1b}$, nm | 428.68 | 445.14 |
| | $a_{1b}$ | $3.59 \times 10^{-4}$ | $3.32 \times 10^{-4}$ |
| | $w_{1b}$ | 157.09 | 169.71 |
| 4 | $\lambda_2$, nm | 599.79 | 588.05 |
| | $a_2$ | $2.08 \times 10^{-4}$ | $1.86 \times 10^{-4}$ |
| | $w_2$ | 104.56 | 97.39 |
| 5 | $\lambda_{2a}$, nm | 594.83 | 593.3888 |
| | $a_{2a}$ | $6. \times 10^{-6}$ | $1.11 \times 10^{-5}$ |
| | $w_{2a}$ | 29.06 | 30.8090 |
| 6 | $\lambda_{3a}$ nm | 643.78 | 640.6494 |
| | $a_{3a}$ | $4.61 \times 10^{-5}$ | $6.45 \times 10^{-5}$ |
| | $w_{3a}$ | 50.74 | 49.96 |
| 7 | $\lambda_3$, nm | 646.68 | 650.27 |
| | $a_3$ | $4.28 \times 10^{-4}$ | $4.099 \times 10^{-4}$ |
| | $w_3$ | 187.93 | 164.2823 |

To understand the physical origin of the constitutive peaks let us remark that water-soluble dyes belong to a wide class of materials called chromonics (for review see [28]). Below we shortly recall physical properties of chromonics that are of interest for the explanation of our results.

Dye molecules as a rule are of a plank-like shape. The main characteristic property of chromonics is the formation of rod-like aggregates, in which dye molecules stack face-to-face onto each other when dissolved in water. The aggregates are poly-disperse in their length (from dimers, threemers, tetramers and so on up to long aggregates such that at higher concentrations they might form liquid crystal phases). If the dye molecules stack onto each other without in-plane shift such that the long axis of the aggregates is perpendicular to the molecular plane, they are called *H-aggregates*. The aggregates, in which each next molecule within the aggregate is shifted with respect to its neighbors such that the long axis of the aggregate is tilted with respect to the molecular planes, are called *J-aggregates*. *H*- and *J*-aggregates differ by their optical spectral properties. *H*-aggregation is accompanied by a shift of light absorption maximum towards shorter wavelength (blue shift) with respect to the non-aggregated molecules (monomers), while *J*-aggregates display red shift. This spectral shift can be as large as from few nanometers to tens of nanometers and therefore can be at work when explaining the position of the absorption band maximum for the system hemoglobin-dye. We did not find in the literature any information concerning the possibility for Cibacron Blue molecules associate into aggregates in water solutions. Nevertheless some estimation can be done basing on the information concerning the aggregation of similar organic polyaromatic dyes. It has been established [29,30] that the dye aggregation is promoted by a high ratio of relative molecular mass ($M_R$) of the dye ionic residue to the total mass ($M_I$) of the ionic groups and the aggregation number can be estimated as

$$N_A = A \times C_{dye} + \frac{M_R}{M_I}, \qquad (6)$$

where $A$ is a constant and $C_{dye}$ is the dye concentration in water. For Cibacron Blue we find $\frac{M_R}{M_I} \approx 2.5$, which is in the range of the values typical for polyaromatic dyes. For example, according to the reference [31] for C.I.Reactive Black dye this value is 2.46 and the $A$ constant is 0.0035 for $C_{dye}$ measured in [$mg/L$]. Taking this value of the $A$ constant for $C_{dye} = 1000 \frac{mg}{L}$ we obtain for Cibacron Blue the aggregation number $N_A \sim 6$. Therefore, the influence of the aggregation of dye molecules on their optical spectral properties can play important role and below we examine it.

Cibacron Blue has the absorption maximum roughly at $\lambda = 615\,nm$ and hemoglobin absorption maximum is at $\lambda = 556\,nm$. Therefore, the observed red shift of the absorption maximum for the system hemoglobin-dye towards longer wavelength (from approximately 615 nm for water dye solutions to approximately 650 nm for hemoglobin-dye mixtures (Figure 5,a)) has to be prescribed to the dye-hemoglobin complexes. It was suggested in [32,33] that the red shift of the absorption band is characteristic for complexation of Cibacron Blue with proteins containing super-secondary structure termed the "dinucleotide fold with a hydrophobic pocket", which the dye molecules occupy in the protein. However our analysis presented above combined with the same Gaussian deconvolution of absorption spectra for pure Cibacron solution without hemoglobin (see below) shows that the apparent red shift of the absorption band is actually due to the redistribution of the absorption weights of the constitutive peaks, rather than due to their shift along the wavelength axis.

We have measured spectra of Cibacron Blue dissolved in water for several different concentrations and performed deconvolution of the absorption spectra into constitutive Gaussian peaks. The shapes of the spectra suggest at least two constitutive peaks: one at short wavelengths below $400\,\text{nm}$ and another one roughly at $615\,\text{nm}$. The result is that two peaks do not fit the experimental data. Then we introduce a third constitutive peak and perform fitting with the following guess positions for the constitutive peaks: $\lambda_1^0 = 400\,\text{nm}$, $\lambda_2^0 = 615\,\text{nm}$, $\lambda_3^0 = 650\,\text{nm}$. The best fitting curve (thick solid line in Figure 5,b) with the best fitting parameters given in Table IV, well coincide with the experimental data (dots in Figure 5,b). The constitutive Gaussian peaks centered at $\lambda_2 = 605\,\text{nm}$ and $\lambda_3 = 649\,\text{nm}$ are plotted by thin lines in Figure 5,b. Second peak at $\lambda_2 = 605\,\text{nm}$ appears to be more intensive than the third one at $\lambda_3 = 649\,\text{nm}$. We have found that positions of the second and third constitutive peaks for different concentrations remain almost the same, though with a weak trend towards longer wavelengths when the dye concentration increases (Figure 5,c).

Table IV. Fitting results of absorption spectra for 0.01 wt.% of Cibacron Blue dye solutions in water (Figure 5,b).

| Peak number | Fitting parameter | Cibacron Blue, 0.01% |
|---|---|---|
| 1 | $\lambda_1$, nm | 151.5 |
| | $a_1$ | 38983669 |
| | $w_1$ | 184 |
| 2 | $\lambda_2$, nm | 605 |
| | $a_2$ | 1467962 |
| | $w_2$ | 110 |
| 3 | $\lambda_3$, nm | 649 |
| | $a_3$ | 119393 |
| | $w_3$ | 46 |

The second peak is much higher than the third one. The ratio $a_2/a_3$ is approximately constant in the studied concentration range (Figure 5,d) and, therefore, these peaks do not display progressive aggregation (desegregation) when the dye concentration increases (decreases) and do not allow to conclude about the type (*H* or *J*) of aggregation. Similar situation was recently reported in the reference [34]. These results show that the determination of the type of aggregation (*H* or *J*) in chromonics via concentration dependencies of absorption spectra is not always conclusive. To determine the aggregation type one needs at least two solutions, in one of which the dye molecules are dissolved in solvent as *separate molecules* without aggregation, whereas in another one they should be in the *aggregated state*. Often this is a serious difficulty, because the aggregation of chromonics exhibits so-called isodesmic behavior [35]: there is no critical concentration at which the aggregation takes place; molecules aggregate even in very diluted solutions. Notice that the isodesmic behavior is agreed with empirical equation (6), which implies that the dye molecules aggregate even when the dye concentration vanish. Moreover for very diluted solutions the detected adsorption signal drops and might become lower than the spectrophotometer noise. In such a situation determination of the type of aggregation via absorption spectra cannot be performed.

Position of the constitutive peaks for pure Cibacron solutions (Table IV, Figure 5b,c) is close to those for hemoglobin-dye solutions (Table III, Figure 5,a): $\lambda_2$-peak of the absorption spectra of pure dye solutions corresponds to $\lambda_2$- and $\lambda_{2a}$- peaks for the hemoglobin-dye solutions, while $\lambda_3$-peak of the dye solution corresponds to $\lambda_3$- and $\lambda_{3a}$- peaks of the hemoglobin-dye solutions. On can suppose that due to binding of the dye molecules to hemoglobin $\lambda_2$- and $\lambda_3$- peaks obtained for pure dye solution split into $\lambda_2$-, $\lambda_{2a}$- and $\lambda_3$-, $\lambda_{3a}$- peaks respectively. Taking into account that $a_{2a} \ll a_2$ and $a_{3a} \ll a_3$ let us compare the ratio $a_2/a_3$ for dye solutions without and with hemoglobin. From Tables III and IV we find that for pure dye solution $\frac{a_2}{a_3} \sim 15$, while for dye with hemoglobin $\frac{a_2}{a_3} \sim 2$. Therefore the apparent red shift of the absorption band after the binding of dye molecules to hemoglobin is a result of significant redistribution of the weights of the absorption peaks: for hemoglobin-dye solution in comparison with corresponding peaks for dye without hemoglobin the $\lambda_2$-peak increases in comparison with $\lambda_3$-peak, while their positions are modified little.

The fact that the $\lambda_3$-peak increases at the expense the $\lambda_2$-peak implies that these peaks are not simply of chemical nature due to the interaction of the dye molecule with the hydrophobic pocket of the hemoglobin molecule as it is suggested in [32,33], but rather have a structural origin. Under structural origin of the peaks here we understand that one of the two peaks $\lambda_2$ or $\lambda_3$ might correspond to the light absorption of the dye monomers, while another one can correspond to the aggregated dye molecules. In other words we suggest that apparent red shift of the adsorption maximum (the shift is apparent because actually it is due to the redistribution of the absorption weights between the constitutive peaks) results from the influence of the protein molecules on the aggregation of dye molecules. In water solution dye molecules are associated into aggregates and this latter implies a shift of absorption maximum for the solution with aggregates with respect to that for non-aggregated molecules towards shorter or longer wavelengths depending on the type (*H*- or *J*-) of aggregation. At least two scenarios can be expected at the binding of dye molecules with the protein molecules: 1) dye molecules can be nipped off from an aggregate and attached to the protein coil *one-by-one* or 2) the aggregate can be attached to the protein molecule *as whole*. In other words, binding of dye molecules with protein according to first scenario will be a process inverse to the aggregation, i.e. the dissociation or disaggregating of the dye molecules and should be accompanied by appearance of a new peak shifted with respect to that for the aggregation or to the change of the ratio of their absorption weights. The superposition of the redistributed peaks will manifests in the spectra as the shift of the experimentally measured absorption band along the wavelength axis. If one can identify the constitutive peaks corresponding to the monomeric and aggregated dye molecules respectively then their relative position along the light wavelength axis will indicate the type of aggregation and the first scenario should show up as the increase of the peak assigned to the monomers accompanied by the decrease of the peak corresponding to the aggregated molecules. If the attaching of the dye aggregates *as whole* to the protein molecules favors the aggregation then one expects the opposite situation: the decrease of the monomeric absorption peak accompanied by the increase of the peak assigned to the aggregates.

Returning to the analysis of the plots for light absorption in the system hemoglobin-dye we state that the same position of the constitutive peaks and the decrease of the $\lambda_2$-peak accompanied by the increase of $\lambda_3$-peak in comparison with the dye solution without hemoglobin, indicates that these spectral changes agree with the idea about the influence of protein on the aggregation of the dye molecules. If binding of dye molecules to hemoglobin is accompanied by the desegregation of dye molecules then one has to accept that the $\lambda_2$-peak corresponds to the aggregates, while the $\lambda_3$-peak is due to monomeric dye molecules and, hence, Cibacron Blue molecules in water are aggregated into *H*-aggregates. Although our opinion (supported by the preliminary spectral absorption data obtained for non-absorbing proteins) is that the binding of dye molecules to proteins splits the dye aggregates, at present we do not have enough data to identify unambiguously the peaks responsible for aggregates and monomers. This latter remains an open question as well as the number of binding sites and the dissociation constant need to be determined. Exchanging the assignment of the $\lambda_2$ and $\lambda_3$-peaks to monomers and aggregates will lead to the opposite conclusion, namely that the Cibacron Blue aggregates are of *J* type. Whether or not but the redistribution of the weights of the constitutive peaks at their almost constant positions after binding dye to hemoglobin can be due to the influence of protein molecules on dye aggregation and, thus, the idea that protein molecules can split dye aggregates into separated molecules through their binding can be useful for the investigations of chromonics to determine the type of aggregation (*H* or *J*) of dye molecules. This test supposes to be more conclusive for non-absorbing proteins, which bind dye molecules, because a shift of the absorption maximum for dye solution mixed with protein would indicate the contribution of the disaggregated dye molecules to the total light absorption.

Another important conclusion following from the analysis of Figure 5 is that hemoglobin extracted from alcoholised rats has better affinity to dye molecules than that for non-alcoholised rats. Indeed, the constitutive peak at $\lambda_{1b} = 428\,\text{nm}$, which is characteristic for pure hemoglobin in water, is present in the spectra of hemoglobin with dye for both *AF* and *AI$_{1,0}$* groups and this indicates that not all the hemoglobin sites responsible for light absorption at $\lambda_{1b}$ are occupied by dye molecules. Thus, the fact that the absorption weight value $a_{1b} = 3.59 \times 10^{-4}$ obtained for the *AF* group, is higher than the corresponding value for the AI$_{1,0}$ group, namely $a_{1b} = 3.32 \times 10^{-4}$ (see Table III and Figure 5), illustrates the conclusion made above according to which hemoglobin molecules extracted from alcoholised rats have better affinity to dye in comparison to the non-alcoholised *AF* group. Affinity of the dye molecules to proteins is of the electrostatic nature [36]. If dye residues are charged opposite to the protein residue, they can be bound to the protein ionic residues. Better affinity implies that the protein ionic residues in $AI_{i,j}$ samples are less densely packed. In other words it means that in the protein molecules, which have been affected by alcohol, some of the links between molecular fragments are broken and, as a result, dye have an access to additional sites in the protein molecule.

Results presented above suggest that ethanol intoxication modifies the structure of blood protein molecules, in particular hemoglobin. A mechanism of these modifications can be based on the reaction with acetaldehyde or active metabolites of oxygen in excess amounts after ethanol consumption.

### 3.CONCLUSION

Our results on comparative spectroscopic studies of blood components extracted from rats intoxicated by ethanol and those free from alcohol intake indicate that ethanol and its metabolites induce conformational modifications of blood proteins. We find that blood of rats intoxicated by ethanol is enriched by the desoxy-form of hemoglobin in comparison with that for alcohol free rats. It is established that alcohol intake during first four months leads to the decrease of fractional weight of oxyhemoglobin and to the increase of methemoglobin amount in blood. Further alcohol consumption is accompanied by recovering of the normal level of hemoglobin derivatives in blood. Normalization of the fractional weight of hemoglobin derivatives in blood after durable (longer than 5-6 months) ethanol intoxication is due to the activation of the enzyme (acetaldehyd dehydrogenase) system, lowering the level of acetaldehydes in blood. Most probably it is at this stage when the mechanisms of adaptation to the alcohol intake activates. Conformational modifications of blood proteins induced by ethanol consumption can be visualized in optical spectra mixing blood protein samples with dyes. Better dye affinity of hemoglobin extracted from alcoholised rats with respect to those from non-alcolised ones confirms that ethanol and its methabolites induces structural pathologies in hemoglobin molecules. The detected changes for the case of the posterity of intoxicated animals may be explained as a post-translation modification, as well as a disturbance of the structure and function of tissue cellular gene mechanism for the blood creation.

**ACKNOWLEDGEMENT**

We acknowledge the Ukrainian State Foundation for the Basic Researches, (the Project N 501) for financial support of this work.

**REFERENCES**

1. U. Berggren, C. Fahlke, E. Aronsson, A. Karanti, M. Eriksson, K. Blennow, D. Thelle, H. Zetterberg, and J. Balldin, "The TaqIA DRD2 A1 allele is associated with

alcohol-dependence although its effect size is small," Alcohol and Alcoholism.**41**(5), 479-485 (2006)

2. H. Tønnesen, L. Hejberg, S. Frobenius, J. Andersen. "Erythrocyte mean cell volume-correlation to drinking pattern in heavy alcoholics," Acta Med. Scand.**219** (5), 515-518 (1986).

3. P.C. Sharpe, R. McBride and G.P. Archbold, "Biochemical markers of alcohol abuse," QJM: An International Journal of Medicine.89, 137-144 (1996)

4. Subir Kumar Das, Prasunpriya Nayak and D.M. Vasudevan, "Biochemical markers for alcohol consumption," Ind.J.Clin.Bioch.**18**(2), 111-118 (2003).

5. A.G. Petrov, "The lyotropic state of Matter," Gordon and Breach, New York, (1999).

6. A. Helamder, B. Tabakoff and Who/Isbra study centers, "Biochemical markers of alcohol use and abuse experiences from the pilot study of the Who/Isbra collaborative project on state and trait markers of alcohol," Alcohol and Alcoholism 32, 133-144 (1997).

7. J.J. Potter, O.A. Mac Dougald, E. Mozey, "Regulation of rat alcohol-dehydrogenase by cyclic AMP in primary hepatocyte culture," Arch. Biochem. Biophys. 321, 329-335 (1995).

8. E.L. Abel, "Alcohol-induced changes in blood gases, glucose, and lactate in pregnant and nonpregnant rats," Alcohol. 13, 281-285 (1996).

9. M.F. Perutz, "Regulation of oxygen-affinity of hemoglobin – influence of structure of the globin on the heme iron," Ann. Review of Biochem. 48, 327-386 (1979).

10. L.B. Ngnuyen, C.M. Peterson, "The effect of acetaldehyde concentrations on the relative rates of formation of acetaldehyde-modified hemoglobins," Proc.Soc.Exp.Biol. 177, 226 (1984).

11. Yu.V. Burov, A.N. Vedernikova, *"Neurochemistry and pharmacology of alcoholism,"* Medicine, Moscow, (1985).

12. K.P. Dudok, O. Moroz, T. Dudok, I. Vlokh, R. Vlokh, "Spectroscopic Study of Haemoglobin Ligand Forms and Erythrocyte Membrane Dynamics at Alcohol Intoxication of White Rats," Ukr.J.Phys.Opt. 5, 32-35 (2004).

13. R.D. Kanitz, W.G. Wood, T. Wetterling, J. Forster, G. Oehler, "New state markers for alcoholism. Comparison of carbohydrate deficient transferrin (CDT) and alcohol mediated (triantennary) transferrin (AMT)," Prog. Neuropsychopharmacol Biol Psychiatry. **18**(3) 431-446 (1994).

14. C.M. Schnitzler, L. Menashe, C.G. Sutton and M.B.E. Sweet, "Alcohol abuse in patients with femoral neck and intertrochanteric fractures," Alcohol and alcoholism. 23, 127-132 (1988).

15. C.K. Farren and K.F. Tipton, "Trait markers for alcoholism: clinical utility," Alcohol and Alcoholism. 34, 649-665 (1999).

16. H.H. Stassen, H. Begleiter, B. Porjesz, J. Rice, C. Scharfetter, T. Reich, "Structural decomposition of genetic diversity in families with alcoholism," Genet Epidemiol. **17**(1), S325-330 (1999).

17. J. Struppe and R.R. Vold, "Dilute Bicellar Solutions for Structural NMR Work," J.Magn.Res. 135, 541–546 (1998).

18. R.S. Prosser, J.A. Losonczi, and I.V. Shiyanovskaya, "Use of a Novel Aqueous Liquid Crystalline Medium for High-Resolution NMR of Macromolecules in Solution," J. Am. Chem. Soc. 120, 11010-11011 (1998).

19. S. Gaemers and A. Bax, "Morphology of three lyotropic liquid crystalline biological NMR media studied by translational diffusion anisotropy," J. Am. Chem. Soc. 123, 12343-12352 (2001).

20. M. Ru1ckert and G. Otting, "Alignment of biological macromolecules in novel nonionic liquid crystalline media for NMR experiments2000, J. Am. Chem. Soc. 122, 7793-7797 (2000).

21. L.G. Barrientos, C. Dolan & A.M. Gronenborn, "Characterization of surfactant liquid crystal phases suitable for molecular alignment and measurement of dipolar couplings," Journal of Biomolecular NMR, 16, 329–337 (2000).

22. L.D. Lukyanova, B.S. Balmukhanov, L.T. Ugolev, *"The oxygen dependence processes in the cell and its functional state,"* Nauka. Moscow, (1982).

23. S. Prahl, "Optical Absorption of Hemoglobin," Oregon Medical Laser Center (1999) [Internet] Available from http://omlc.ogi.edu/spectra/hemoglobin/index.html

24. J.E. Hale, V. Gelfanova, J.R. Ludwig and M.D. Knierman, "Application of proteomics for discovery of protein biomarkers," Brif.Fun.Gen.Prot. 2, 185–193 (2003).

25. S.T. Thomson ST, K.H. Cass & E. Stellwagen, "Blue Dextran-Sepharose: An Affinity Column for the Dinucleotide Fold in Proteins," Proc.Nat.Acad.Sci. U.S.A. 72 669-672 (1975)

26. M.G. Rossman, A. Liljas, C.I. Branden & L.J. Banaszak, "Evolutionary and structural relationships among dehydrogenases," in *The Enzymes,* P.D. Boyer, 3rd Eds., pp.61-102, Academic Press, New York, (1975).

27. J-F. Bielmann, J-P. Samana, C.I. Branden and H. Eklund, "X-Ray studies of the binding of Cibacron blue F3GA to liver alcohol dehydrogenase," Eur. J. Biochem. 102, 107-110 (1979).

28. J. Lydon, "Chromonics," in *Handbook of Liquid Crystals*, D. Demus, J. Goodby, G.W. Gray, H-W. Spiess and V. Vill, Eds., pp. 981–1007, Wiley-VCH, Weinheim, (1998).

29. J. Shore, "*Cellulosics Dyeing*," Society of Dyers and Colourists, Bradford (1995).

30. A. Datyner, A.G. Flowers and M.T. Pailthorpe, "A study of dyestuff aggregation. I. Determination of dye particle sizes by light scattering," J.Coll.Interf.Sci. 74, 71–79 (1980).

31. L. Shu, T.D. Waite, P.J. Bliss, A. Fane, V. Jegatheesan, "Nanofiltration for the possible reuse of water and recovery of sodium chloride salt from textile effluent," Desalination 172, 235–243 (2005).

32. G.E. Schulz and R.H. Schirmer, "Topological comparison of adenyl kinase with other proteins," Nature 250, 142-144 (1974).

33. M.G. Rossman MG, Moras D, and K.W. Olsen,"Chemical and biological evolution of nucleotide binding protein," Nature 250, 194-199 (1974).

34. Yu.A. Nastishin, H. Liu, T. Schneider, V. Nazarenko, R. Vasyuta, S.V. Shiyanovskii and O.D. Lavrentovich, "Optical characterization of the nematic lyotropic chromonic liquid crystals: Light absorption, birefringence, and scalar order parameter," Phys.Rev.E. 72 041711, 1-14 (2005).

35. P.K. Maiti, Y. Lansac, M.A. Glaser and N.A. Clark, "Isodesmic self-assembly in lyotropic chromonic systems," Liquid Crystals. 29, 619–626 (2002).

36. B. Salih and R. Zenobi, "MALDI mass spectrometry of dye-peptide and dye protein complexe," Anal. Chem. 70, 1536-1543 (1998)